\documentclass[british,12pt, a4paper,twoside,openright]{article} 
\usepackage[T1]{fontenc}
\usepackage[utf8]{inputenc}
\usepackage{color}
\usepackage{latexsym}
\usepackage{indentfirst}
\usepackage{graphicx}
\usepackage{amsmath,amssymb,amsfonts,amsthm,amstext,amscd}
\usepackage{mathrsfs}
\usepackage{fancyhdr}
\usepackage{times}
\usepackage{enumerate}
\usepackage{natbib}
\usepackage[hang,flushmargin]{footmisc}

\makeatletter
\newtheorem{defi}{Definition}[section]

\newtheorem{lem}{Lemma}[section]
\newtheorem{rem}{Remark}[section]

\newtheorem{coro}{Corolary}[section]

\fancypagestyle{plain}{%
\fancyhf{}%
\fancyhead[LO]{\textbf{Science \& Philosophy}
}
\fancyhead[RO]{Volume 9(2), 2021, pp. xx-xx   }
\fancyfoot[C]{\thepage }

\fancyfoot[LE,RO]{ \vspace{1cm} doi:xxxx-xxxxx-xxx-x  }
}

\usepackage{url}
\usepackage{breakurl}
\usepackage{stmaryrd}

\makeatother

\usepackage{babel}

\begin{document}
\title{\textbf{On the relation of free bodies, inertial sets and arbitrariness}}
\author{Hernán Gustavo Solari\footnote{Departamento de Física, FCEN-UBA and IFIBA-CONICET;
Pabellón I, Ciudad Universitaria (1428) - C.A.B.A - Argentina. email:
solari@df.uba.ar Orcid: 0000-0003-4287-1878}  {\ }and
             Mario Alberto Natiello\footnote{Centre for Mathematical Sciences, Lund University.
Box 118, S 221 00 LUND, Sweden. email: mario.natiello@math.lth.se
(corresponding author) Orcid: 0000-0002-9481-7454}}

\date{}

\maketitle

\begin{abstract}
\begin{normalsize}
\noindent
We present a fully relational definition of inertial systems based
in the No Arbitrariness Principle, that eliminates the need for absolute
inertial frames of reference or distinguished reference systems as
the ``fixed stars'' in order to formulate Newtonian mechanics. The
historical roots of this approach to mechanics are discussed as well.
The work is based in part in the constructivist perspective of space
advanced by Piaget. We argue that inertial systems admit approximations
and that what is of practical use are precisely such approximations.
We finally discuss a slightly larger class of systems that we call
``relatively inertial'' which are the fundamental systems in a relational
view of mechanics.\\

\footnote{\noindent Received on October 3, 2021. Accepted on December 31, 2021.
  Published on zzz,zzz. doi: 10.23756/sp.v8i2.5xx. ISSN 2282-7757; eISSN 2282-7765.
  \copyright The Authors. This paper is published under the CC-BY licence agreement.}
\end{normalsize}
\end{abstract}

\newpage

\section{Introduction}

In the wording of Newton, ``absolute time'', ``absolute space''
and ``absolute motion'' \citep[Escholium to Definition VIII]{newt87}
are the ingredients of concern in the formulation of the laws of mechanics.
These ideas led to a core concept of Newtonian mechanics, namely that
of \emph{inertial frame}, i.e., the frame(s) on which free bodies
move with constant velocity (or where changes in the velocity of a
body correspond to forces/interactions). However, Newton's concept
of absolute space rests on multiple grounds, some of them fundamental
for his theory of mechanics, others more adapted to his philosophical
taste while not mandatory for the laws of motion.

Newton's philosophical position was challenged by Leibniz who questioned
the ``absolute'' character of space on philosophical and theological
grounds, proposing that space is not absolute and prior to the objects
we observe, but it is indeed given by the relations among objects,
a standpoint that came to be called \emph{relationism}. Other relationist
positions were later held by C. Neumann, Streintz, Mach, Thomson and
Langer \citep{disa90}, although arriving to different conclusions.
Neumann and Streintz assumed the existence of a privileged frame (the
'Alpha Body' in Neumann and 'any free body' in Streinz, \citep{disa90})
while Mach questioned absolute space as a metaphysical construct.
Thomson and Langer, on the other hand, attempted to construct inertial
frames based on the motion of free bodies.

The notion of absolute space in Newton presents several aspects (for
a recent discussion based in the work by H. Stein, see \citep{disa20})
but not all of them are \emph{operational}, i.e., have consequences
in the formulation and explanatory power of the theory. For Newton:
\begin{defi}
\emph{Absolute space} \label{par:First-definition}
\end{defi}
\begin{quote}
Absolute space, in its own nature, without regard to anything external,
remains always similar and immovable. Relative space is some movable
dimension or measure of the absolute spaces; which our senses determine
by its position to bodies; and which is vulgarly taken for immovable
space;{[}...{]}
\end{quote}
and later, further discussing about the properties of space he writes:
\begin{defi}
\emph{True motion} \label{par:Second-definition}
\end{defi}
\begin{quote}
The causes by which true and relative motions are distinguished, one
from the other, are the forces impressed upon bodies to generate motion.
True motion is neither generated nor altered, but by some force impressed
upon the body moved; but relative motion may be generated or altered
without any force impressed upon the body. \citep{newt87}
\end{quote}
It is true motion what is operational in Mechanics and furthermore,
there is no deductive reasoning that brings us from absolute space
into true motion and much less, there is a form to logically infer
the first notion from the acceptance of the second. In Newton's belief
absolute space is where absolute (or true) motion takes place (``Absolute
motion is the translation of a body from one absolute place into another;{[}...{]}''
\citep{newt87}).

The relationist position sustains that there are only spatial relations
between bodies, and consequently, space is not more than a construction
that allows to present such relations, easily connecting with our
intuitions. Piaget \citep[Ch. 1]{piag99} observed that basic notions
of space emerge during the first month of childhood simultaneously
with the notion of object and the discovery of ``permanence'' (and
then change) as well as groups of operations. The relationist position
was presented by Leibniz in his exchange with Clarke, who defended
Newton's view. The centre of the exchange between Clarke and Leibniz
is precisely what goes from Definition \ref{par:First-definition}
to Definition \ref{par:Second-definition} and the necessity, as well
as correctness, of the first one. No dispute emerges regarding the
second notion. Leibniz initiates the discussion by objecting
\begin{quote}
Newton says that space is an organ--- · like a sense-organ · ---
by which God senses things. But if God needs an organ to sense things
by, it follows that they don’t depend entirely on him and weren’t
produced by him.\citep[first letter]{leibniz2007}
\end{quote}
a theological argument indeed. By the fifth exchange, the objection
had evolved into an epistemic one:
\begin{quote}
\textquotedbl I answer that indeed motion doesn’t depend on being
· observed, but it does depend on being · observable. When there is
no \emph{observable} change there is no motion---indeed there is
no change of any kind. The contrary opinion is based on the assumption
of \emph{real absolute space}, and I have conclusive refuted that
through the principle of the need for a sufficient reason.\textquotedbl{}
\citep[fifth letter]{leibniz2007}
\end{quote}
Leibniz' positioning is epistemic, and concerns a sort of ``hygiene
of reason'' that can later be found in Peirce:
\begin{quote}
A hypothesis is something which looks as if it might be true and were
true, and which is capable of verification or refutation by comparison
with facts. The best hypothesis, in the sense of the one most recommending
itself to the inquirer, is the one which can be the most readily refuted
if it is false. \citep[CP 1.120]{peir94}
\end{quote}
and later
\begin{quote}
Long before I first classed abduction as an inference it was recognised
by logicians that the operation of adopting an explanatory hypothesis
--which is just what abduction is-- was subject to certain conditions.
Namely, the hypothesis cannot be admitted, even as a hypothesis, unless
it be supposed that it would account for the facts or some of them.
\citep[CP 5.189]{peir94}
\end{quote}
The first notion of absolute space \eqref{par:First-definition} does
not admit experimental refutations since nothing is logically deduced
from it. The totality of Newton's arguments such as the famous ``bucket''
\footnote{Search for ``vessel'' in \citep{newt87}}\citep{barb82}
rests upon the insight on absolute space given by ``true motion'',
Definition \ref{par:Second-definition}.

Thomson went one step further in the discussion arguing that the second
form can be deduced from spatial relations between objects and he
named ``inertial frames'' the systems providing the references that
fulfil the requirement. In this form, the ``sensorium of God'' \citep{leibniz2007}
is deemed not only epistemologically improper but unnecessary as well.

A better known line of argumentation was entertained by Mach:
\begin{quote}
Newton's experiment with the rotating vessel of water simply informs
us, that the relative rotation of the water with respect to the sides
of the vessel produces \emph{no} noticeable centrifugal forces, but
that such forces \emph{are} produced by its relative rotation with
respect to the mass of the earth and the other celestial bodies. \citep[p. 232]{mach19}
\end{quote}
Later he returns to the point with his famous view with regard to
the fixed stars, while confronting Streintz about Newton's distinction
between absolute and relative rotation (that Streintz accepts):
\begin{quote}
For me, only relative motions exist {[}...{]}, and I can see, in this
regard, no distinction between rotation and translation. When a body
moves relatively to the fixed stars, centrifugal forces are produced;
when it moves relatively to some different body, and not relatively
to the fixed stars, no centrifugal forces are produced. I have no
objection to calling the first rotation \textquotedbl absolute\textquotedbl{}
rotation, if it be remembered that nothing is meant by such a designation
except \emph{relative rotation} \emph{with respect to the fixed stars}.
\citep[p. 543]{mach19}
\end{quote}
It has been already indicated that Mach in this respect misses the
point \citep{borzeszkowski1995}. Let us explain the matter with a
simple example that may be reproduced at home. Consider first the
Olympic sport of ``hammer throw''. To make the throw the athlete
makes the hammer rotate around a vertical axis that goes approximately
from his feet to a point in the line from his body to the hammer.
The hammer can be said to rotate around the vertical axis and the
athlete rotates around the same axis. To play the sport strong muscles
are needed to pull from the hammer producing the internal force needed
to rotate. Newton would call this absolute rotation because force
and motion are in correspondence. Let the hammer rest on the floor
and spin around your feet. Observe how all objects around appear to
rotate (in a geometrical sense) relative to you. Despite some of the
objects being far more massive than the hammer, they rotate almost
effortless. Newton calls this apparent or relative rotation as it
lacks correspondence between force and geometrical description. Since
reality cannot be reduced to geometrical appearance we have to distinguish
both situations. As indicated in \citep{borzeszkowski1995}, Mach's
difficulties are rooted in his philosophical standing as empiricist.\footnote{The idea that natural science rests on experience is not controversial.
What is controversial is the form in which the observed (or sensed)
becomes an idea, the process of ideation. Most empiricists consider
the observed as the fact and the fact as the real, thus effectively
suppressing the subject, they become realists (objective) by the unconscious
action of ignoring the process of production of ideas. For other philosophers
such as \citet[CP 5.145][]{peir94}, there is an ideated reality and
an observable reality; they bear the relation that exists between
universal and particular. Experience is experience of the particular
(uniquely situated in time and space an in several other forms) and
cannot be used in any other case. This particular experience fosters
and tests explanatory ideas, conjectures; by the processes of abduction
experiences nourish the real (ideated) which is always transitory,
a temporary belief. By the process of interpretation and contrast
(induction in the words of Peirce) we control the validity of our
beliefs. For Peirce then, the process of abduction is and inexcusable
part of science.\citet{mcauliffe2015}} It is interesting to recall that Newton had rejected the idea of
the absolute space being in correspondence with fixed stars.
\begin{quote}
... For it may be that there is no body really at rest, to which the
places and motions of other may be referred.

But we may distinguish rest and motion, absolute and relative, one
from the other by their properties, causes and effects. It is a property
of rest, that bodies really at rest do rest in respect to one another.
And therefore, as it is possible that in remote regions of the fixed
stars, or perhaps far beyond them, there may be some body absolutely
at rest; but impossible to know, from the position of bodies to one
another in our regions whether any of these do keep the same position
to that remote body; it follows that absolute rest cannot be determined
from the position of bodies in our regions. \citep{newt87}
\end{quote}
It is important to understand at this point that Newton's and Mach's
views have a point in common: they want to resolve the problem of
inertia by introducing a unique, universal, conjectured reference
system be it explicitly metaphysical or not. In contrast, Thomson
will depart from this view by considering that references are to be
found within what is experimentally at disposition which is indeed
the actual form in which we use Newton's laws to explain observed
phenomena. Think for example of a table-top experiment demonstrating
2d-collisions. The following demonstration was presented by Prof.
Miguel Ángel Virasoro \footnote{https://en.wikipedia.org/wiki/Miguel\_Ángel\_Virasoro\_(physicist)}
to his students: the table was made of well polished marble and was
well levelled; the carts consisted of a cylindrical base made of steel
with a recipient above it where frozen $CO_{2}$ was placed. The $CO_{2}$
gas (produced by sublimation) flew out of the recipient between the
base and the marble, lifting the cart just above the table reducing
friction substantially. The reference system provided by the table
may be (approximately) inertial or not. In addition to this real demonstration
we have to imagine the possibility of giving the table a rotatory
motion. Being almost decoupled from the carts, the rotation of the
table does not change the interactions between carts. Yet, it changes
the description of the trajectories. We are in the presence of apparent
rotation. To confuse matters more, the room where the table is could
be rotating as well, so that the inertial system would be given by
a table rotating with respect to the room. But yet, in any case, in
the idealised friction-less situation, there will always exist an
inertial system which is not deductible from appearances.\footnote{The position of Mach as relationist cannot be distinguished from that
of Neumann, the Alpha Body of Neumann being the distant stars of Mach.
This is most evident when Mach discusses inertial mass and he states
\begin{quote}
\emph{Definition of equal masses}: All those bodies are bodies of
equal mass, which, mutually acting on each other, produce in each
other equal and opposite accelerations.\citep[p.218]{mach19}
\end{quote}
Unless we define a reference system, there is no such thing as an
acceleration for \emph{each} body. Considering two bodies, only the
relative distance makes sense, and how the relative acceleration is
distributed among the bodies is not known. By adequately choosing
the acceleration of the frame of reference we can make the allocation
the way it pleases us. Thus, the definition does not make sense unless
we specify the reference system as the ``distant stars'' or we explicit
are speaking of ``true motion'' in Newton's sense. Inertial systems
are discussed in Mach in the appendix, p.545\textcolor{black}{, while
commenting work of Lange that appeared after the first edition of
Mach's book.} Thus, despite his claims regarding being a relationist,
essential parts of Mach's thoughts are based in a secular version
of Newton's absolute space. It is interesting to realise that the
context of the discussion are the various attempts to reformulate
Newton's mechanics such as \citet{hertz89}, \citet{streintz1883}
and \citet{lange1886} (see \citep{Pfister2014} as well) trying in
one way or another to dispose of the metaphysical absolute space.}

Thomson opens his paper with:
\begin{quote}
There is no distinction known to men among states of existence of
a body which can give reason for any state being regarded as a state
of absolute rest in space, and any other being regarded as a state
of uniform motion... the only motion of a point that men can know
of or can deal with is motion relative to one, two, three bodies or
more other points. \citep[p. 568-569]{thomson1884}
\end{quote}
For Thomson, and later for Lange, it is clear that the notion of space
must be sought in the relation between bodies (``...qualities or
distinctions of motions of one or more bodies can be ascertained through
knowable relations between these motions and the motions of one or
more other bodies...'' \citep[p. 572]{thomson1884}). The reference
of motion corresponds to the concept of ``free bodies'', he states:
\begin{quote}
The only motion of a point that men can know of or can deal with is
motion relative to one, two, three, or more other points {[}...{]}
Any arrangement whatever of points, lines, or planes, changeless in
mutual configuration, will, for present purposes, be named as a reference
frame, or briefly as a \emph{frame}.\citep[p.569-560]{thomson1884}
\end{quote}
In the bi-dimensional space provided by the marble-table of the suggested
demonstration, the bodies move freely as long as they do not collide
with each other (or with the contention rails that prevent them from
falling). The motion of such bodies, relative to each other will give
the references for any other motion. It is then wise to adjust the
table so that the free bodies move in straight lines at constant speed
with respect to it. In such a form, the table has acquired the same
property than free bodies and we can use it as a reference frame in
Thomson's view.

A free body is certainly a concept, an idea, that emerges by the process
of idealisation introduced by \citet[day 4]{gali14}. We can say that:
\begin{defi}
\emph{\label{AppFree} Approximately free}: A body is said to be approximately
free when for a given purpose its interactions with other bodies can
be ignored within the established level of tolerance.
\end{defi}
There are several instances in which a body can be considered approximately
free because its characteristics (such as charge, mass, volume and
others), its relative position to other bodies, and the lapse of time
that is relevant for our purpose, allow us to expect no appreciable
influence from other bodies on it or, what it is for practical purposes
the same, influences are cancelled (as it is usually the case for
electrical influences between statistically neutral bodies).

\begin{defi}
\emph{\label{Def-Free-body} Free body}: A free body is the idealised
version of an approximately free body.
\end{defi}
Thus, free bodies exist only in our mind and, at the same time, approximately
free bodies are not too difficult to find if we exercise some due
tolerance. Further, Thomson proposed that inertial reference frames
can be introduced by considering the motion of bodies with respect
to a reference point and directions derived from the relative motion
of free bodies.

In \citep{sola18b}, notions of space and time were developed elaborating
from the concepts of \emph{me }vs.\emph{ other }and\emph{ permanence}
vs. \emph{change}. Starting from the subjective intuitive notion of
space we proposed the existence of \emph{inertial frames}, and a relational
formulation of Newtonian mechanics implying that absolute space (in
the sense objected by Leibniz) was not fundamental to Newton's theory.
The development in \citep{sola18b} rests on the No Arbitrariness
Principle (NAP), namely that \emph{no knowledge of Nature depends
on arbitrary decisions.} In other words, in any description of Nature
arbitrariness is either absent or ``controllable'', in the sense
that different arbitrary representations are connected by a \emph{group}
of transformations.

In the present work we connect, in a mathematical form, the notion
of a relational space with (a) the intuitions developed during the
early childhood \citep{piag99}, (b) the notion of intrinsic reference
systems as developed by Thomson and (c) the No Arbitrariness Principle.
As a result, we show that the class of reference frames that preserve
the objective relational dynamics includes non-rotating ``accelerated''
systems and that only a subclass of them consists of Thomson's inertial
frames.Neither the fixed stars or the absolute space of Definition
\ref{par:First-definition}are required (or relevant) to set the grounds
of Newtonian mechanics. The historical approach is then reverted,
from ``free bodies move in straight uniform motion with respect to
an inertial frame'' into ``inertial frames are those reference systems
where free bodies are described as moving in straight uniform motion''.
Since the idea of ``free, non-interacting, body'' admits approximations,
inertial systems admit approximations as well. Actually, the intuitive
use of inertial frames corresponds to this later notion.

In the next Section we start by developing the notion of space from
a Piagetian perspective, then connecting it with the Cartesian concept
of coordinates. Since Descartes presentation assumes the existence
of an external observer, we identify the relational content of the
notion of space and finally develop the theory of relatively inertial
frames, and its connection to Newtonian Mechanics. We discuss the
achievements in the final Section.

\section{The relational notion of space\label{sec:The-perception-of}}

Piaget describes the construction of the notion of space, objects
and spatial relations in the child:
\begin{quote}
To understand how the budding intelligence constructs the external
world, we must first ask whether the child, in its first months of
life, conceives and perceives things as we do, as objects that have
substance, that are permanent and of constant dimensions. If this
is not the case, it is then necessary to explain how the idea of an
object (object concept) is built up. The problem is closely connected
with that of space. A world without objects would not present the
character of spatial homogeneity and of coherence in displacements
that marks our universe. Inversely, the absence of “ groups” in the
changes of position would be equivalent to endless transformations,
that is, continuous changes of states in the absence of any permanent
object. In this first chapter, then, substance and space should be
considered simultaneously, and it is only through abstraction that
we shall limit ourselves to object concept.\citet[p. 3]{piag99}

The conclusion to which the analysis of object concept has led us
is that in the course of his first twelve to eighteen months the child
proceeds from a sort of initial practical solipsism to the construction
of a universe which includes himself as an element. At first the object
is nothing more, in effect, than the sensory image at the disposal
of acts; it merely extends the activity of the subject and, without
being conceived as created by the action itself (since the subject
knows nothing of himself at this level of his perception of the world),
it is only felt and perceived as linked with the most immediate and
subjective data of sensorimotor activity. During the first months
the object does not, therefore, exist apart from the action, and the
action alone confers upon it the quality of constancy. At the other
extreme, on the contrary, the object is envisaged as a permanent substance
independent of the activity of the self, which the action rediscovers
provided it submits to certain external laws. Furthermore, the subject
no longer occupies the center of the world, a center all the more
limited because the child is unaware of this perspective; he places
himself as an object among other objects and so becomes an integral
part of the universe he has constructed by freeing himself of his
personal perspective. \citet[p. 97]{piag99}
\end{quote}
To address the notion of velocity, first we need to address the notion
of spatial relations. Take e.g., a look at a garden. Leave aside (project
out) the moving leaves and birds, and consider those elements that
impress us as keeping a constant relation among themselves (stones,
bushes, etc.). We seek for a universal instruction to move around
the garden. We then identify the elements with labels, $i\in\left[1\dots N\right]$
and summarise the instruction of ``going from $i$ to $j$'' as
$x_{ij}$. Any moving instruction can be given as a concatenation
of moving instructions, this is the most essential condition of spatial
relations. We use the symbol $\oplus$ to denote concatenation. $x_{jj}$
denotes the instruction for remaining at the locus of $j$, or just
``do nothing''. We call $x_{jj}$ the \emph{neutral} element, $0\equiv x_{jj}$.
We then realise that $x_{ij}=x_{ij}\oplus x_{jj}$, and further
\begin{eqnarray}
x_{ij}\oplus x_{jk} & = & x_{ik}\nonumber \\
x_{ij}\oplus x_{jj} & = & x_{ij}\label{eq:feno-space}\\
x_{jj} & \equiv & 0\nonumber \\
x_{ij} & = & \ominus x_{ji}\nonumber 
\end{eqnarray}
 The last line in \eqref{eq:feno-space} express the perceived fact
that the outcome of staying in one place in the end is the same as
the concatenation of going from that place to any other and returning.
Thus, \emph{returning} becomes the inverse operation of \emph{going}:
$x_{ij}=\ominus x_{ji}$.

We need now to introduce velocities, and hence, we first need to introduce
change, or its abstract form, time (that time is the abstract form
of change was already known to Aristotle \citep{aris50bce}). Our
observations may indicate/suggest that the organisation of the garden
is not always the same, perhaps because we want to explain where a
bird is feeding in the garden. We have then decided that there are
things that, for our purposes, are permanent (do not change) such
as trunks and stones as well as changing objects, e.g., the position
of birds. ``Going to bird 1'' is not the same type of instruction
as ``going to tree 1''. The birds cannot be located with ``old
instructions'', the location instructions have to be updated as a
function of other perceived changes. Each observer may have her/his
own clock, so we write $t_{S}$ to record the changes perceived by
$S$\footnote{Notice that clocks in general come from outside the phenomena in study,
they are the remaining part of the universe from which our mind is
isolating the observed. The perception that some events are recurrent
forces us to avoid arbitrary decisions by assigning to the intervals
between consecutive events the same duration. In turn, each interval
can be divided using other series of events (more frequent) equally
perceived as having the same essence, thus refining the division of
time. The repetition and idealisation of the process gives us the
time.}, and understand that all the instructions in Eq. \ref{eq:feno-space}
were given for a determined time, namely:
\begin{eqnarray*}
x_{ij}(t_{S})\oplus x_{jk}(t_{S}) & = & x_{ik}(t_{S})\\
x_{ij}(t_{S})\oplus x_{jj}(t_{S}) & = & x_{ij}(t_{S})\\
x_{jj}(t_{S}) & \equiv & 0\\
x_{ij}(t_{S}) & = & \ominus x_{ji}(t_{S})
\end{eqnarray*}
Since we have an intuition of regular processes, we may agree on the
functioning of reference clocks and hence consider time-records as
real numbers with the usual sum operation. We finally consider the
relation between the rate of change for our (reference) clock and
the rate of change of relative positions, 
\[
v_{ij}(t_{S},\delta)\equiv\frac{x_{ij}(t_{S}+\delta)\ominus x_{ij}(t_{S})}{(t_{S}+\delta)-t_{S}}
\]
 It follows that
\begin{equation}
v_{ij}(t_{S},\delta)\oplus v_{jk}(t_{S},\delta)=v_{ik}(t_{S},\delta)\label{eq:rel-vel}
\end{equation}
This is, whatever the clock is, the composition law between velocities
must be the same that the composition law of the space.

\subsection{Descartes' mathematisation of space}

The Cartesian view is always the view of an observer, the view that
matches our intuitive construction, namely an extrinsic view. In Descartes'
method, directions and distances are used instead of giving instructions
to move around the garden based upon landmarks. Thus, the instruction
that was ``walk from $i$ to $j$, $x_{ij}$'' becomes ``from $i$
walk $x$ steps in the direction $\hat{e}_{ij}$ to $j$'', that
we annotate $x_{ij}=\hat{e}_{ij}x$. If we now agree to consider only
the path from a given reference position (the position of ``ego''),
all paths consist in concatenations of this kind of instructions.
Our intuition tell us more, it tells us that there are only three
independent directions, at least as much as we can perceive. Therefore,
the space is three dimensional and the mathematical construct is Cartesian
space, represented by $\mathbb{\mathbb{R}}^{3}$, while $\oplus$
is the ordinary vector sum. Cartesian space not only inherits the
rules developed for the concatenation of instructions, but adds new
rules based on intuition, such as $x\hat{e}+x^{\prime}\hat{e}^{\prime}=x^{\prime}\hat{e}^{\prime}+x\hat{e}$
(addition is commutative), as well as the other rules of vector algebra.
This idea is underlying the operational part of Newton's concepts
of absolute space and true motion, seeking for a reference --somehow
external to the process in study-- from which changes in motion correspond
to forces.

\subsection{Subjective and relational spaces \label{subsec:Subjective-and-relational}}

A central issue in Classical Mechanics has been that of formulating
mechanics in a fully relational way, this is, not only suppressing
the absolute space but suppressing a Cartesian view based upon a privileged
reference system as well. Absolute space can be seen as the view of
a privileged observer: God. Nothing is gained by changing the name
of the privileged observer. Already with Leibniz the alternative idea
of a relational space arose, i.e., a space free of the arbitrariness
of an extrinsic reference. How do we construct an intrinsic view (i.e.,
without external observers)?

Individual subjective spaces contain the arbitrariness of the choice
of origin and the choice of references, but what is \emph{real} presents
``characters that are entirely independent of our opinions about
them'' \citep[p. 18]{peir55}. In this sense, the Cartesian space
is not \emph{real}. In contrast, relational constructions as those
we used to introduce this discussion, like $x_{ij}$, stand their
chance of being real. Yet, the Cartesian view is not completely arbitrary
because all arbitrary spaces that we can produce map into each other
in a one-to-one form. Thus, the observations in one space only need
to be translated into the observations in another space (characterised
by different arbitrariness). We say that the descriptions are \emph{intersubjective}.
When the differences between subjective spaces correspond to arbitrary
elections that influence the description in a systematic form, as
it is the case of the choice of origin and the choice of directions
of reference, the set of transformations relating the different descriptions
must satisfy conditions of consistency that allow us to move in the
set of arbitrary descriptions without contradictions. This is the
core meaning of the No Arbitrariness Principle (NAP) \citep{sola18b},
in short: the set of transformations associated to arbitrary decisions
must form a \emph{group}. Actually, considering it in finer detail,
there is a group associated with each class of equivalent arbitrary
decisions, this is: a group for the election of reference point, a
group for the election of reference directions, and so on.

We begin by considering one body alone in a $3$-dimensional, universe.
For such a body, relative space makes no sense at all. There is nothing
else available to consider (such as e.g., relative positions), apart
from the body. When we consider two bodies, only a one dimensional
universe is conceivable. The distance between the two bodies is the
only possibility for geometric change. When we consider at least three
bodies, a distinct difference arises. Arbitrariness in the representation
corresponds to the choice of different orientations and the location
of one point in the system. If the Cartesian space for $N$ bodies
corresponds to $R^{3N}$and the group of transformations between arbitrary
representations (after restricting the choice of directions to orthogonal
directions) is $\mathbb{E}(3)=\mathbb{ISO}(3)=\mathbb{SO}(3)\ltimes\mathbb{R}^{3}$
(in words: $\mathbb{E}(3)$ equals the semi-direct product of its
subgroups $\mathbb{SO}(3)$ and $\mathbb{R}^{3}$ - normal subgroup-,
hence $\mathbb{SO}(3)\cong\mathbb{E}(3)/\mathbb{R}^{3}$). The two
component groups correspond to a global orientation and the position
of one point. The real space is what results of modding out the arbitrariness.
This means: a point for $N=1$, the positive line for $N=2$. For
$N\ge3$ it acquires the characteristics we intuitively assign to
the relational space by removing from the subjective space $R^{3N}$
e.g., a global orientation and a distinguishable point.

\subsection{Relatively inertial frames}

The goal of this Section is to define a reference frame internal to
a set of $N$ bodies moving without (relative) interactions, i.e.,
$N$ free bodies.
\begin{lem}
\label{Lemma-motion}The law of motion for the relative distance $x_{ik}$
among any pair of free bodies is 
\[
\frac{dx_{ik}}{dt}=C_{ik},
\]
 for some constants $C_{ik}$.
\end{lem}
\textbf{Proof. }  %\begin{proof}
According to NAP, the law of motion for the relative position, $x_{ik}$,
must be independent of the existence of other bodies since they cannot
influence the motion (otherwise, the pair is not free or it is influenced
by arbitrary decisions regarding the other bodies). Hence, 
\[
\frac{dx_{ik}}{dt}=f(x_{ik})=f(x_{ij})+f(x_{jk})=f(x_{ij}+x_{jk}).
\]
Therefore, $f$ must be an affine transformation, $f(x_{ik})=C_{ik}+Ax_{ik}$,
with $C_{ik},A$ constants. We have the additional result that $C_{ik}=C_{ij}+C_{jk}$,
a principle of addition of the velocities of free bodies. In addition,
since the law must be the same for all times (there is no privileged
time), $C_{ik},A$ must be constant. Finally, $A=0$ since otherwise
the bodies would not be independent of each other, being their evolution
affected by their relative distance.$\Box$

Let $i,j\in(1,2,3)$. We have the relations $x_{ij}$ (oriented distances)
and we can consider a large number of different vectors, e.g. the
set $\lbrace x_{ij},\,dx_{ij}=x_{ij}(t_{S}+\delta)-x_{ij}(t_{S})\rbrace$.
In a three dimensional space those vectors are not completely arbitrary:
some internal relations will become explicit. Three bodies define
a plane, which along with the relative velocity vectors \ref{AppFree}
to define a reference frame, except under singular (full coplanarity
or zero velocity) circumstances. We are now in the position to consider
\emph{relatively inertial} systems.

We set
\begin{equation}
\hat{e}_{ij}=\begin{cases}
\frac{x_{ij}}{|x_{ij}|} & if\,\frac{d}{dt_{S}}x_{ij}=0\\
\frac{v_{ij}}{|v_{ij}|} & if\,\frac{d}{dt_{S}}x_{ij}\equiv v_{ij}\ne0
\end{cases}\label{eq:rel-dir}
\end{equation}
where $v_{ij}=\lim_{\delta\rightarrow0}v_{ij}(t_{S},\delta)$. The
use of a limit is not strictly necessary, since we will deal with
constant $v_{ij}$. It is enough to ask for $v_{ij}$ being independent
of $\delta$.

In terms of an arbitrary frame, the conditions $dx_{ij}=0$ and $dv_{ij}=0$
are ultimately a perception of the observer and as such they introduce
subjectivity in the description. It is this decision made by the subject
what creates the space that can be mathematically represented. Eq.
\ref{eq:rel-dir} removes much of the subjectivity leaving only the
arbitrary origin of the subjective frame, which is not involved in
the description of relative positions. Thus, the observer might be
linearly accelerated but the description of relative motion will remain
unchanged.
\begin{defi}
\label{def:ReIn} Two bodies are \emph{relatively inertial} if there
exists a reference frame, a constant vector $a$\emph{, }and a scalar\emph{
$b$ }such that ${\displaystyle \frac{db}{dt_{S}}}$ is constant and
\begin{eqnarray*}
\hat{e}_{ij}\times\left(x_{ij}\times\hat{e}_{ij}\right) & = & a\\
\hat{e}_{ij}\cdot x_{ij} & = & b\\
\frac{d}{dt_{S}}\hat{e}_{ij} & = & 0
\end{eqnarray*}
where
\[
x_{ij}=a+b\hat{e}_{ij}
\]
\end{defi}
\begin{defi}
\label{N3}$N\ge3$ bodies are relatively inertial if there exists
a reference frame and a sequential order $1,\cdots,N$ such that body
$k$ is relatively inertial to body $k+1$. We call this frame a \emph{relatively
inertial frame} for the $N$ bodies.
\end{defi}
\begin{coro}
Free bodies are relatively inertial bodies.
\end{coro}
\textbf{Proof. }  %\begin{proof}
This is a consequence of Definition \ref{Def-Free-body}, Definition
\ref{N3} and Lemma \ref{Lemma-motion}.
$\Box$ %\end{proof}
\begin{lem}
Relatively inertial is an equivalence relation i.e., a relation $\mathcal{\sim}$
such that for three bodies $A,B,C$ it holds that $A\mathcal{\sim}A$,
$A\mathcal{\sim}B\Rightarrow B\mathcal{\sim}A$, $A\mathcal{\sim}B\,\mathrm{and}\,B\mathcal{\sim}C\Rightarrow A\mathcal{\sim}C$.
\end{lem}
\textbf{Proof. }  %\begin{proof}
$A\sim A$ since whenever $x_{ii}=0$ and for arbitrary $\hat{e}$
we have $a=0$ and $b=0$ in Definition \ref{def:ReIn}. If a pair
$a,b$ exists such that $A\sim B$ it follows that $B\sim A$ with
the associated pair $-a,b$. Since there is a common reference frame
for the whole chain of relatively inertial bodies, the third relation
follows from vector addition rules. We have
\begin{eqnarray*}
x_{AB} & = & a+b\hat{e}_{AB}\\
x_{BC} & = & a^{\prime}+b^{\prime}\hat{e}_{BC}\\
x_{AC} & = & x_{AB}+x_{BC}\\
 & = & a+a^{\prime}+b\hat{e}_{AB}+b^{\prime}\hat{e}_{BC}
\end{eqnarray*}
Further, $b\hat{e}_{AB}+b^{\prime}\hat{e}_{BC}=c+t_{S}\left(\frac{db}{dt}\hat{e}_{AB}+\frac{db^{\prime}}{dt}\hat{e}_{BC}\right)$,
where $c$ is some constant vector and the constant quantity in parenthesis
is either zero or some other constant nonzero vector. In the latter
case, letting $\lambda=||\frac{db}{dt}\hat{e}_{AB}+\frac{db^{\prime}}{dt}\hat{e}_{BC}||$
and $\hat{e}_{AC}={\displaystyle \frac{1}{\lambda}\left(\frac{db}{dt_{S}}\hat{e}_{AB}+\frac{db^{\prime}}{dt}\hat{e}_{BC}\right)}$
we get $x_{AC}=\left(a+a^{\prime}+c\right)+\lambda t_{S}\hat{e}_{AC}$.
$\Box$   %\end{proof}
\begin{lem}
\label{lem:eq.class} The relatively inertial frames referring to
$N\ge3$ relatively inertial bodies have as group of arbitrariness
the translations (which may be time dependent) and the (time independent)
rotations composed as a semi-direct product group.
\end{lem}
\textbf{Proof. }  %\begin{proof}
The instantaneous relative position of $N$ bodies is invariant under
$\mathbb{ISO}(3)$ as explained in Subsection \ref{subsec:Subjective-and-relational}.
With relative positions, the arbitrariness of the origin of coordinates
cancels out, even when it changes as a function of time. In contrast,
a change in the arbitrary choice of orthogonal directions of reference
as a function of time will make $\frac{d}{dt_{S}}\hat{e}_{ij}\ne0$
in the definition \ref{def:ReIn}, hence it will break the concept
of relatively inertial set.
$\Box$  %\end{proof}
We observe that in the cited paragraphs of Thomson, the directions
of reference in space correspond to the relative motion of free bodies.
Relatively inertial frames correspond to the situation presented in
\citet{newt87}, Collorary VI of the chapter Axioms. The introduction
of relatively inertial sets of bodies deserves a detailed discussion.
In subsection \ref{subsec:Subjective-and-relational} we introduced
the instantaneous space for the description of the relational problem
of $N$ bodies. Starting from subjective space, $R^{3N}$, we arrived
to a relational space where the orbits of points by the action of
the group $\mathbb{E}(3)=\mathbb{ISO}(3)$ were identified, this is:
$R^{3N}/\mathbb{E}(3)$. The description of the evolution in time
in the relational space is then represented by a function of time,
$R$, into $R^{3N}/\mathbb{E}(3)$. Thus, each trajectory is given
by a function $F(t):R\mapsto\left(R^{3N}/\mathbb{E}(3)\right)$ and
the set of functions will be called ${\cal F}$. Definition \ref{def:ReIn}
uses relative positions, $x_{ij}$, which are invariant under changes
of the origin of the (subjective) coordinates, hence, these mathematical
objects are invariant under the action of $\mathbb{R}(3)$, the group
of translations, and only rotations in $\mathbb{E}(3)$ may change
them. From the original group of arbitrariness, $\mathbb{E}(3)$,
we are left with the effective action of $\mathbb{E}(3)/\mathbb{R}(3)\sim\mathbb{SO}(3)$
(a result that can be intuited as well). Since we have to make such
a choice for every time when considering a trajectory, the group of
arbitrariness associated to the set $\left\{ x_{ij}(t)\right\} $
consists of (continuous and twice differentiable) time-dependent rotations.
Continuity and differentiability is requested because we have to deal
with velocities in the definition. Because the translation of the
subjective origin of coordinates does not intervene in the definition
of the inertial set and inertial frame, we can allow any arbitrariness
to such a collective translation. Note that the arbitrariness of the
description encompasses $\mathcal{F}$ and time-dependent rotations,
while the class of relatively inertial frames related to a relatively
inertial set of bodies only contains time-independent rotations, because
of Lemma \ref{lem:eq.class}.

Regarded from any arbitrary relatively inertial frame, the $N$ relatively
inertial bodies belonging to the relatively inertial system may display
any possible type of trajectory $x_{i}(t)$ as indicated above. However,
for any pair of bodies we have that $x_{i}(t)-x_{j}(t)=v_{ij}t+x_{ij}(0)$,
with constant $v_{ij}$, when the reference frame is chosen with orientations
as indicated in definition \ref{def:ReIn}. Picking a reference frame
that is also relatively inertial to the $N$ bodies, the standard
description of the bodies from an inertial frame \citep[ch. 1, ][]{gold80}
is recovered. Indeed, the $N$ bodies are then described as having
coordinates $x_{i}(t)=x_{i}(0)+v_{i}t$. This view is equivalent to
augmenting the set of bodies in one, adding an extra ``body'', representing
the origin of coordinates. The transformations among inertial frames
in the standard setting, the Galilean group, arise as a consequence
of this choice (it corresponds to picking different extra bodies relatively
inertial to the set of $N$, having different constant relative velocity).
\begin{defi}
\emph{\label{Inertial-frame} Inertial frame}. Inertial frames are
the relatively inertial frames associated to free bodies.
\end{defi}
\begin{coro}
Newton's laws hold only in inertial frames.
\end{coro}
It is possible for a set of bodies to be relatively inertial without
being free, hence two separate sets of bodies can be each relatively
inertial and yet no inertial frame may be available for all the bodies
to pertain to a relatively inertial set.
\begin{rem}
The ``added value'' of the concept of relatively inertial is twofold.
On one side, the concept is intrinsic to a system and hence independent
of the system being e.g., accelerated relative to some external reference.
Moreover, it is fully compatible with classical, Newtonian mechanics,
thus eliminating a metaphysical ingredient discussed for centuries,
now without any reference to e.g., ``distant stars''. In an inertial
system, any relative motion that deviates from constant velocity corresponds
with forces/interactions, i.e., to bodies that are not free. However,
the statement is not true if we replace ``inertial system'' by ``relatively
inertial system''\footnote{In \citep{sola18b} it is assumed explicitly that the reference corresponds
to a free body.}.
\end{rem}
\begin{rem}
The relatively inertial class is larger that the inertial class associated
to the inertial frames \ref{Inertial-frame}. It contains any kind
of global time-dependent displacement of the frame and the objects
under study and, in particular, accelerated systems as those considered
in Einstein's equivalence principle (stating the complete physical
equivalence of a homogeneous gravitational field and a corresponding
acceleration of the reference system \citep{einstein1907relativity}).
Regarded in this way, the equivalence principle is not an independent
principle but a particular case of the more general NAP and the presently
derived concept of relatively inertial\@.
\end{rem}
\begin{rem}
The concept of force in Classical Mechanics has been gradually ``naturalised''
by habit (in the sense that forces became treated as observables,
rather than as meta-observables --ideated--). Consequently, an inertial-mass
has followed, relating force and the actually observable acceleration.
However, a detailed construction of mechanics \citep{sola18b} shows
that the concept of inertial mass rests completely on the concept
of gravitational mass. Therefore, it makes no sense to distinguish
them. Hence, the ``weak equivalence principle'' \citep{dick81}
(stating that inertial mass equals gravitational mass) is a social
construction rather than a basic ingredient of mechanics.
\end{rem}

\section{Discussion and conclusions}

It is worth to discuss the implications of Definition \ref{AppFree}.
In practice, we always deal with approximate notions, be it of free
bodies, relatively inertial bodies or inertial frames. The possibility
of interacting or not with other objects/bodies is always considered
from a limited perspective and it is subject to revision whenever
necessary. While perfection is probably unattainable outside the mathematical
universe, the pair of concepts \emph{approximation} and \emph{tolerance}
are quantifiable and admit the possibility of refinement and improvement.

We have shown that there are two different conceptions of relational
space, one that preserves the intuition of the subjective space and
constitutes a secular version of absolute space resting upon hypothetical
fixed stars, and a second one, almost forgotten, that rests upon the
concept of inertial frames, free bodies and internal relations. Such
distinction parallels the notion of Absolute space (Definition \ref{par:First-definition})
in Newton and that of absolute motion (Definition \ref{par:Second-definition})
which is the operative concept in Newton's \emph{Principia}. The first
one can be called metaphysical as well as it is not true that for
any practical use we identify inertial systems by referring them to
the fixed stars or any other equivalent primary reference as Neumann's
Alpha Body or the distribution of masses of a --necessarily assumed--
finite universe. The second version is of practical use but does not
support the simple intuition as crystallised in the Cartesian representation
of space, but rather requires a philosophical intuition \citet{huss83}
or equivalently, to acknowledge the process of abduction in general
and the production of reality by the building intelligence of the
child.

Leibniz' relationism is the result of a discipline of mind based on
the ``Principle Sufficient Reason'' and the ``Principle of Identity
of Indiscernibles'' which relate deeply with the ``No Arbitrariness
Principle'' and the process of modding out arbitrariness (subjective
views) applied in the present construction. In contrast, providing
an arbitrary reference for the space, such as fixed stars, violates
NAP, and as such accepts Leibniz' conclusions regarding the space
but disdains Leibniz' main rational principles.

In terms of classical mechanics we have shown that it is possible
to introduce inertial frames and sets of inertial bodies without introducing
an Absolute Space (or universal references of no actual use) complementing
the work in \citep{sola18b}. We have also shown that the new approach
includes a larger variety of reference systems.

In the view of the present work, Newton's definition \ref{par:First-definition}
of absolute space plays only a metaphysical role. There is no way
of working out from absolute space the meaning of true motion (definition
\ref{par:Second-definition}) and there is no way to put absolute
space to test, as Newton explains in the Scholium. We suggest that
its need must be considered in psychological terms as it relates to
initial perceptions of the child in the construction of the notion
of space. Absolute space can be safely replaced with a secular relational
space such as the one produced by taking the ``fixed stars'' as
alleged reference, since we are only changing the useless metaphysics.
Why should free bodies be in uniform motion in the frame of reference
given by the fixed stars? Yet, the metaphysical discussion has entertained
most philosophers \citep{barb82}.

In contrast, the notion of true motion (definition \ref{par:Second-definition}),
closely linked to the principle of inertia, can be inferred from observations
and the abducted result has stood firmly all falsification attempts.
It stands as a true belief, one that ``shapes our actions'' \citep[CP 5.371]{peir94},
on its own right. In this work we have shown how inertial frames follow
logically from it. Since inertial frames correspond to the setting
where Newton's axioms hold true, hence all of Newton's mechanics rests
on the notion of true motion.

In achieving our result we rested, much as Leibniz did, on a sanitising
concept, in our case the ``No Arbitrariness Principle'' (NAP) and
its mathematical formulation. While NAP operates before Newton's laws
of mechanics, a residual action is left. Since there is not a unique
inertial system, there should be a group of operations transforming
statements in one inertial system into statements in another one,
while the laws of mechanics remain the same. Thus, this relativity
principle is not an independent construct somehow unique to physics
but only a consequence of NAP, having to account for a residual arbitrariness.

\section{Acknowledgements}

MAN acknowledges grants from Kungliga Fysiografiska Sällskapet 2016
and 2018.

\bibliographystyle{plainnat}

\end{document}